\documentstyle [12pt]{article}
\begin{document}
\thispagestyle{empty}
\author{{Hong-Liang Lu$^{\#}$, and Xi-Jun Qiu}}
\title{{\bf Some Exact Results of Hopfield Neural Networks and
 Applications }
\footnotetext{\# {\bf Hong-Liang Lu}, Corresponding author, Ph.D student
major to
 Radio Physics in Physics Department, Science College. Address: Physics
 Department,
 Science College, 20 Chengzhong Road, Jiading, Shanghai 201800, China.
 Email: xjqiu@srcap.stc.sh.cn\\
\indent\indent\indent{\bf Xi-Jun Qiu}, Professor and Ph.D supervisor of
Theoretical Physics, Physics Department, Shanghai University.}}
\date{{\small{\it Physics Department, Science College,
Shanghai University}}}
\maketitle
 \begin{abstract}
 A set of fixed points of the Hopfield type neural network was under
 investigation. Its connection matrix is constructed with regard to the
 Hebb
 rule from a highly symmetric set of the memorized patterns.
 Depending on the external parameter the analytic description of the
 fixed
 points set had been obtained. And as a conclusion, some exact results
of
 Hopfield neural networks were gained.
 \baselineskip 0.3in
 \\{\bf PACS Number(s):} 87.10.+e, 05.45.+b
 \\{\bf Key Words:} Neural Networks, Hebb rule, Fixed points
 \end{abstract}
 \vskip 0.2in
 \baselineskip 0.3in
 Over the past decade, there has been an explosion of interest in
 so-called artificial neural network(ANN) technology. ANNs are a model
of
 computing inspired by the brain[1]-[5], consisting of a collection of
 model "neurons" connected by model "synapses." Computation in the
 network is
 performed in a distributed fashion, by propagating excitatory and
 inhibitory activations across the synapses, and computing the neuronal
 outputs as
 a nonlinear (typically sigmoidal) function of total synaptic input.
 These
 networks also have a capacity to "learn" to perform a given computation
 by adjusting real-valued synaptic connection strengths (weight values)
 between units. ANNs are of considerable interest both for biological
 modeling of information processing in the nervous system, and for
 solving
 many classes of complex real-world applications.
 
 J.J.Hopfield boosted neural network research at the beginning of the
 1980s with the publication of a famous paper on artificial neural
 networks,
 which he used for pattern completion and to solve optimization
 problems[4]. These networks consist of one layer of neurons that are
 completely
 connected with each other. Hopfield analysed the behavior of networks
 belonging to that type, and could prove mathematically that stable
 behavior
 may be achieved under certain conditions.
 
 It can be shown that the dynamic behavior of Hopfield type neural
 networks is described by an energe surface. Each network state
 corresponds to a
 certain position on that surface. Through external clamping, neurons
may
 be forced to certain states of activity, and thus the whole network may
 be
 forced to move to a well defined point on the energy surface. If the
 network is released, {\it i.e.} external clamping is removed, it will
 change its state
 in such a way that it moves on the energy surface towards new
 states of lower energy. Finally, neuron states will stop changing if a
 local minimum in the energy surface is reached. Through careful
 selection of
 weights, ocillations will be avoided.
 A set of fixed points of the Hopfield type neural network[4][6]is under
 investigation. Its connection matrix is constructed with regard to the
 Hebb
 rule from a $(p\times n)$-matrix $\bf S$ of memorized patterns:
 $${\bf S}=\left(\begin{array}{ccccccc}
 1-x&1&\ldots&1&1&\ldots&1\\
 1&1-x&\ldots&1&1&\ldots&1\\
 \vdots&\vdots&\ddots&\vdots&\vdots&\ldots&\vdots\\
 1&1&\ldots&1-x&1&\ldots&1\end{array}\right).$$
 Here $n$ is the number of neurons, $p$ is the number of memorized
 patterns $\vec s^{(l)}$, which are the rows of the matrix $\bf S$, and
 $x$ is
 an arbitrary real number.
 
 Depending on $x$ the memorized patterns $\vec s^{(l)}$ are interpreted
 as
 $p$ distorted vectors of the {\it standard}
 $$\vec\varepsilon (n)= (\underbrace{1,1,\ldots,1}_n).\eqno(1)$$
 
 We denote by $\vec\varepsilon (k)$ the configuration vector which is
 collinear
 to the bisectrix of the principle orthant {\it standard-vector}. Next,
 $n$ is the number of the spin variables, $p$ is the number of the
 memorized
 patterns and $q=n-p$ is the number of the nondistorted coordinates of
 the standard-vector.
 Configuration vectors are denoted by small Greek letters. We use small
 Latin letters
 to denote vectors whose coordinates are real.
 
 The problem is as follows: {\it the network has to be learned by
 $p$-times
 showing of the standard (1), but a distortion has slipped in the
 learning
 process.
 How does the fixed points set depends on the value of this distortion
 $x$?}
 
 Depending on the distortion parameter $x$ the analytic description of
 the
 fixed points set has been obtained. It turns out to be very important
 that the
 memorized patterns $\vec s^{(l)}$ form a highly symmetric group of
 vectors:
 all of them correlate one with another in the same way:
 $$
 (\vec s^{(l)},\vec s^{(l')})=r(x),  \eqno (2)
 $$
 where $r(x)$ is independent of $l,l'=1,2,\ldots,p.$ Namely this was the
 reason
 to use the words "highly symmetric" in the title.
 
 It is known [7], that the fixed points of a network of our kind have to
 be of
 the form:
 $$\vec\sigma^*=(\sigma_1,\sigma_2,\ldots,\sigma_p,1,\ldots,1),\quad
 \sigma_i=\{\pm 1\},\ i=1,2,\ldots,p.\eqno(3)$$
 Let's join into one {\it class} $\Sigma^{(k)}$ all the {\it
 configuration}
 vectors $\vec\sigma^*$ given by Eq.(3), which
 have $k$ coordinates equal to "--1"
 among the first $p$ coordinates. The class $\Sigma^{(k)}$ consists of
 $C_p^k$ configuration vectors of the form (3), and there are $p+1$
 different classes $(k=0,1,\ldots,p)$. Our main result can be formulated
 as a
 Theorem.
 \\{\bf Theorem.} {\it As $x$ varies from $-\infty$ to $\infty$ the
fixed
 points set is exhausted in consecutive order by the classes of the
 vectors[8]
 $$\Sigma^{(0)},\Sigma^{(1)},\ldots,\Sigma^{(K)},$$
 and the transformation of the fixed points set from the class
 $\Sigma^{(k-1)}$
 into the class $\Sigma^{(k)}$ occurs when $x=x_k$:
 $$x_k= p\frac{n-(2k-1)}{n+p-2(2k-1)},\quad k=1,2,\ldots,K.$$
 If $\frac{p-1}{n-1}<\frac13$, according this scheme all the $p$
 transformations
 of the fixed points set are realized one after another and $K=p$. If
 $\frac{p-1}{n-1}>\frac13 $, the transformation related to
 $$K=\left[\frac{n+p+2}4\right]$$
 is the last. The network has no other fixed points.}
 
 The Theorem makes it possible to solve a number of practical problems.
 We would
 like to add that the Theorem can be generalized onto the case
 of arbitrary vector
 $$\vec u =(u_1,u_2,\ldots,u_p,1,\ldots,1),\quad \sum_{i=1}^p u_i^2=p$$
 being a standard instead the standard (1). Here memorized patterns
 $\vec s^{(l)}$ are
 obtained by the distortion of the first $p$ coordinates of the vector
 $\vec u$
 with regard to the fulfillment of Eqs.(2).
 
 The obtained results can be interpreted in terms of neural networks,
 Ising model and factor analysis.
 \vskip 5mm
 {\bf Acknowledgement} The authors acknowledge the benefit of extended
 interaction with Miss Jie-yan Bai of Shanghai Research and Development
 Center for Fiber Optic Technology, Shanghai 803 Research Institute, who
 has helped us develop or clarify several ideas and issues that appear
in
 this paper. We also extend our thanks to Prof. Yu-Long Mo of the School
 of Information and Communication Engineering, Shanghai University.
 

\vskip 0.4in
 \begin{thebibliography}{s40}
 \bibitem{s1}McCulloch M.S. W.Pitts, A logical caculus of the ideas
 immanent in nervous activity, {\it Bull. of Math. Biophys.}{\bf 5},
 1943,pp.115-133
 \bibitem{s2}Hebb D.O., {\it The organization of Behavior}, Wiley, New
 York,1949
 \bibitem{s3}Rosenblatt F., {\it Principles of Neurodynamics},Spartan
 Books, 1962
 \bibitem{s4}Hopfield J.J., Neural networks and physical systems with
 emergent collective computational abilities, {\it Proc. Natl. Acad.
Sci.
 U.S.A.},
 Vol.{\bf 79}, 1982, pp.2554-2558
 \bibitem{s5}Rumelhart D.E., McClelland J.L., {\it Parallel Distributed
 Processing}, Vol.{\bf 1}, {\bf 2}, MIT Press, Cambridge, MA, 1986
 \bibitem{s6}Hopfield J.J., Neurons with Graded Respone have collective
 computational properties like those of two-state neurons, {\it Proc.
 Natl.
 Acad. Sci. U.S.A.}, Vol.{\bf 81}, 1984, pp.3088-3092
 \bibitem{s7}L.B.Litinsky. Direct calculation of the stable points of a
 neural network. {\it Theor. and Math. Phys.}{\bf 101}, 1492 (1994)
 \bibitem{s8}L.B.Litinsky. Fixed points of Hopfiled type neural
networks,
 cond-mat/9901251 (1999)
 \end{thebibliography}
 \end{document}